\documentclass{iaus}
\usepackage{graphicx}

\title[H$_2$ in PN NGC\,6369] 
{Weak and Extended H$_2$ Emission in NGC\,6369}

\author[Ramos-Larios et al.] 
{G.\ Ramos-Larios$^1$, M.A.\ Guerrero$^2$, R.\ V\'azquez$^3$, \and J.P.\ Phillips$^1$\thanks{Deceased, April 
29th, 2011}}

\affiliation{
$^1$Instituto de Astronom\'{\i}a y Meteorolog\'{\i}a, Universidad de Guadalajara, Jalisco, Mexico \\
email: {\tt gerardo@astro.iam.udg.mx} \\

$^2$Instituto de Astrof\'{\i}sica de Andaluc\'{\i}a, IAA-CSIC, Granada, Spain \\
email: {\tt mar@iaa.es} \\ [\affilskip]

$^3$Instituto de Astronom\'{\i}a, Universidad Nacional Aut\'onoma de M\'exico, Ensenada, Mexico\\ 
email: {\tt vazquez@astro.unam.mx} \\[\affilskip]}

\pubyear{2011}
\volume{283}  
\pagerange{1--2}
\jname{Planetary Nebulae. An eye to the future, E3, 1 of 2}
\editors{A.\ Manchado, L.\ Stanghellini \& D.\ Sch\"onberner, eds.}
\begin{document}

\maketitle

\begin{abstract}
NGC\,6369 is a double-shell PN with a filamentary outer shell or envelope and faint bipolar extensions. We have used ground- and space-based narrowband optical and near-IR images, broadband mid-IR images, optical long-slit echelle spectra, and mid-IR spectra to investigate its physical structure. The observations confirm a bipolar structure for the inner shell of NGC\,6369, but they also reveal evidence for H$_2$ and strong polycyclic aromatic hydrocarbons (PAHs) emission from a photo-dissociation region (PDR) with molecular inclusions located outside the bright inner shell. 
 
\keywords{(ISM:) planetary nebulae: individual: NGC\,6369, infrared: ISM}
\end{abstract}

\firstsection

\section{Introduction}

There is a growing number of PNe in the literature showing knots or blobs of emission well separated from their main nebular shells that do not form a halo.  
In some cases, the kinematics of these knots imply high-velocity collimated outflows, e.g., Fleming~1 (\cite[Boffin \& Miszalski 2011]{2011IAUS283}; \cite[Palmer \etal\ 1996]{PLML96}) or MyCn\,18 (\cite[Redman \etal\ 2000]{Redman_etal00}), while in some others, as for IC\,4634 (\cite[Guerrero \etal\ 2008]{Guerrero_etal08}), the velocity of the knots is close to the systemic velocity, may be due to motions close to the plane of the sky.

\begin{figure}[t]
\begin{center}
 \includegraphics[bb=85 0 755 461,width=0.455\columnwidth]{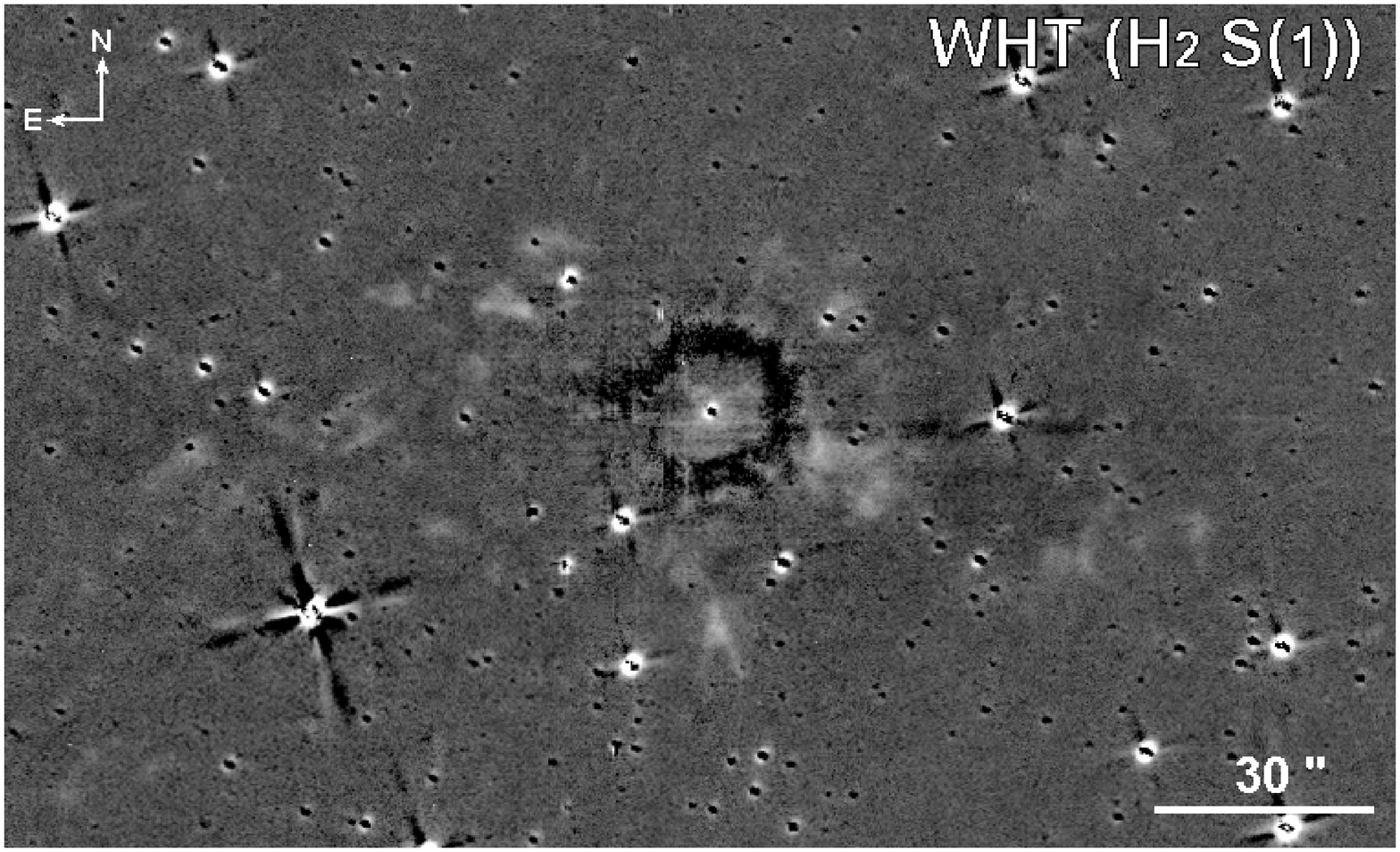}
 \hspace*{0.5\columnsep}%
 \includegraphics[bb=70 268 615 701,width=0.39\columnwidth]{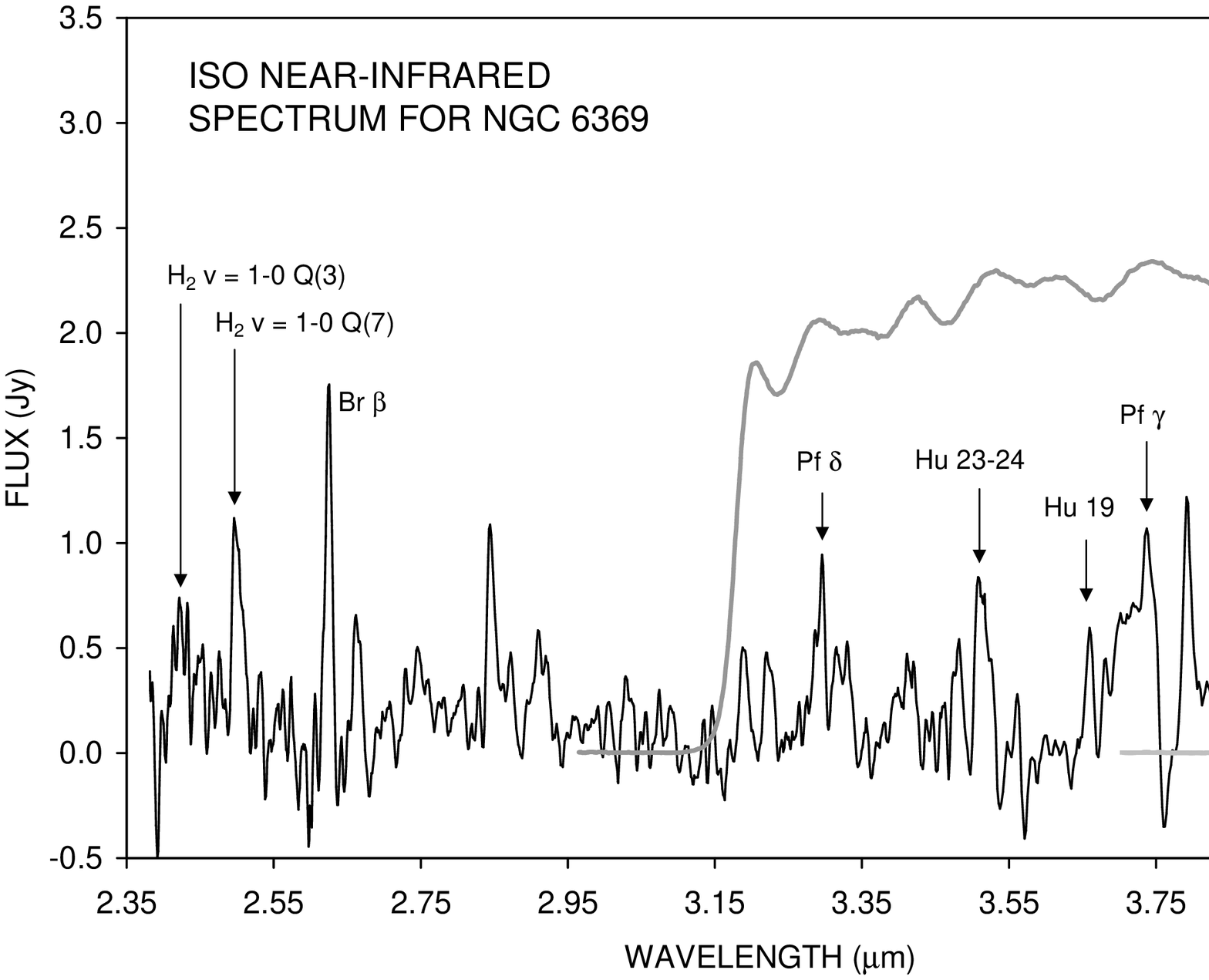}
\caption{({\it left}) LIRIS H$_2$ background-subtracted image of NGC\,6369, and ({\it right}) \emph{ISO} spectrum of NGC\,6369 in the 2.4--4.1 $\mu$m spectral range. The dark and light diffraction spikes in the LIRIS image are due to the rotation of the focal reducer during the image acquisition}
   \label{fig1}  
\end{center}
\end{figure}

\section{Observational Results}

[O~{\sc iii}], H$\alpha$, and [N~{\sc ii}] narrowband images of NGC\,6369 were obtained with ALFOSC at the 2.56m NOT, and near-IR H$_2$ (1,0) S(1), Br$\gamma$, and continuum $K$ narrowband images were obtained with LIRIS at the WHT (Figure\,\ref{fig1} - {\it left}). An inspection of the optical and near-IR images of NGC\,6369 shows a bright inner shell, two extensions, and an envelope. The western extension is reminiscent of a lobe or a large {\it ``ansa''}, whilst the eastern one displays a complex morphology that can be described as a bifurcated structure best seen in the H$\alpha$ and [N~{\sc ii}] images.
The envelope, barely detected in [O~{\sc iii}] and only a little brighter in H$\alpha$, has bright [N~{\sc ii}] emission composed of an ensemble of knots, blobs, and filaments. The region interior to the bright inner shell, which shows anomalously low excitation emission, is revealed to be filled with H$_2$ emission. We found evidence that the [N~{\sc ii}] knots and filaments are associated with H$_2$ emission, i.e., they include significant amounts of neutral material. 
 
\emph{Spitzer} IRAC images were retrieved from the NASA/IPAC Infrared Science Archive (IRSA) obtained as part of Program 20119 (The Darkest Cloud, An IRAC/MIPS Survey of the Pipe Nebula, PI: C.J. Lada). The eastern and western extensions, as well as the outer envelope, show a higher relative contribution of 8.0 $\mu$m emission.

Long-slit high-dispersion spectroscopy of NGC\,6369 has been acquired using MES on the 2.1m OAN-SPM telescope. The spectral 
range includes the H$\alpha$ and [N~{\sc ii}] $\lambda6583$ lines. 
Five long-slit spectra were obtained to map the kinematics at different regions of the nebula with spectral resolution of $\approx$12 km~s$^{-1}$. The position-velocity maps reveal that the envelope is not completely inert.

An \emph{ISO} spectrum of NGC\,6369 in the mid-IR spectral range is showing in Figure\,\ref{fig1} - {\it right}. The \emph{ISO} spectral range is dominated by emission lines, with Br$\gamma$ being the most intense and the 3.3 $\mu$m PAH band being relatively weak or negligible. 
The data also show convincingly that the emission in the IRAC 3.6 and 4.5 $\mu$m bands is largely dominated by H~{\sc i} lines, with the Br$\alpha$ $\lambda$4.052 $\mu$m line in the 4.5 $\mu$m IRAC band being the brightest line.  
The evidence for the H$_2$ 1$-$0 Q(3) $\lambda$2.424 $\mu$m and H$_2$ 1$-$0 Q(7) $\lambda$2.500 $\mu$m lines is consistent with the detection of H$_2$ emission in our WHT LIRIS images and at 17.036 $\mu$m (\cite[Pottasch \& Bernard-Salas 2008]{PB-S08}).  

\section{Discussion}

The inner optical shell in NGC\,6369 is the brightest feature in optical, near-IR and mid-IR images, whereas the outer envelope and east and west extensions are fainter but still noticeable.
We report the detection of irregular knots and blobs of diffuse emission in low-excitation and molecular line emission that are located up to 80$^{\prime\prime}$ from the central star, well outside the main nebular shells. 

These external condensations and filaments associated to the bipolar extensions may probe earlier, point-symmetric ejections, but they can also arise from shocks associated with the east-west bipolar flow that excite the material surrounding the nebula.

\end{document}